\documentclass[12pt,nofootinbib]{revtex4}

\usepackage{amsmath, amssymb, amsfonts, latexsym}
\usepackage[matrix,arrow,curve]{xy}

\newcommand{\oR}{{\mathbb R}}
\newcommand{\oC}{{\mathbb C}}
\newcommand{\oZ}{{\mathbb Z}}

\renewcommand{\Re}{\mathop{\mathrm{Re}}\nolimits}
\renewcommand{\Im}{\mathop{\mathrm{Im}}\nolimits}

\newcommand{\eqdef}{\stackrel{\mathrm{def}}{=}}
\newcommand{\curl}{\mathop{\mathrm{curl}}\nolimits}

\newtheorem{lemma}{Lemma}

\begin{document}

\thispagestyle{empty}

\hbox to \textwidth{\hfil FIAN-TD/2016-18 $\phantom{xxxx}$}
\hbox  to \textwidth{\hfil Phys. Rev. D94  (2016) 105021$\phantom{xxxx}$}

\bigskip

\bigskip

\title{Dirac's monopole, quaternions, and the Zassenhaus formula}

\author{M.~A.~Soloviev}
\email{soloviev@lpi.ru}

\affiliation{
\medskip
\baselineskip=15pt I.~E.~Tamm Department of Theoretical Physics,
P.~N.~Lebedev Physical Institute,\\ Russian Academy of Sciences,
Leninsky Prospect~53,  Moscow 119991, Russia}

\begin{abstract}
\medskip
\baselineskip=15pt
Starting from the quaternionic quantization scheme proposed by Emch and Jadczyk for describing the motion of a quantum particle in the  magnetic monopole field,  we derive
an algorithm for finding the differential representation of the star product generated by the quaternionic Weyl correspondence on  phase-space functions.
This procedure is illustrated by the explicit calculation of the star product up to the second order in the Planck constant $\hbar$. Our main tools are an operator analog of the twisted convolution  and the Zassenhaus formula for the products of exponentials of noncommuting  operators.
\end{abstract}
\maketitle

\baselineskip=20pt

\section{Introduction}\label{S1}

 Since the works by Wu and Yang~\cite{WY1,WY2,WY3}  and   Greub and Petry~\cite{GP}, it has been generally recognized that the theory of fiber bundles provides the most appropriate mathematical framework for  the quantum mechanical description of the motion of a charged particle in the field of the Dirac magnetic monopole. The interpretation of the particle wave function as a section of a vector bundle associated with the Hopf bundle reveals the topological nature of  Dirac's charge quantization condition~\cite{D1,D2} and  provides
 a consistent,  singularity-free formulation overcoming the problem of the absence of a globally defined vector potential for the monopole field. There are numerous papers on this subject, but the fiber-bundle description of the motion of a quantum particle in the monopole field continues to attract attention because it is prototypical for many aspects of quantum gauge theory.  In an interesting paper, Emch and Jadczyk~\cite{EJ} have proposed  a  magnetic monopole model based on using a quaternionic Hilbert space and the   concept of a weak  projective group representation introduced by Adler~\cite{A1,A2,AE}.
 The feasibility of  such an approach stems from the fact that the Hopf fibration $SU(2)\to SU(2)/U(1)=S^2$ is obtained from the   trivial principal fiber bundle  $S^2\times SU(2)\to S^2$ by reducing the structural group from $SU(2)$ to $U(1)$ (see, e.g.,~\cite{Madore} for this reduction).

 An interesting question is how to define a generalized Weyl map which associates quaternionic operators  to phase-space functions  in the Emch and Jadczyk setting. This question was raised  in~Ref.~\cite{CG-BLMV}, where  an attempt was   made to find an integral representation of the noncommutative  star product generated by this map on the phase space. However,  for comparison with the general formulation~\cite{K} of deformation quantization of Poisson manifolds, a differential form of this  product  is needed. In the present paper, we develop a regular procedure for finding the differential form  of the star product, starting from the multiplier of the  quaternionic projective representation~\cite{EJ}  of the group of configuration-space translations, and we calculate this product explicitly up to the second order in the Planck constant~$\hbar$. Our main tools are an operator analog of the twisted convolution used in the usual Weyl calculus~\cite{Mail,Fol,W} and the Zassenhaus formula which is a convenient combinatorial expansion of $e^{X+Y}$ with noncommuting $X$ and $Y$.   A detailed description  of this formula and its relation to the  Baker-Campbell-Hausdorff theorem can be found in Ref.~\cite{CMN}. We also obtain   an exact expression for the integral kernel of the star product.
 This result agrees well with the composition rule~\cite{M,MP} for  gauge-invariant Weyl symbols, which was derived for a  quantum particle in a  magnetic field having a global vector potential $\mathbf A(x)$, although  the derivation method used in Refs.~\cite{M,MP} and based on the replacement of the canonical momentum $P$ in the Weyl system by the kinetic momentum $P-(e/c)\mathbf A(x)$ is inapplicable to the magnetic monopole case.

 The paper is organized as follows.  In Sec.~\ref{S2}, we recall the Hamiltonian description of the motion of a charged particle in a magnetic field, which uses a noncanonical Poisson structure  depending on the magnetic field. In Sec.~\ref{S3}, we consider the quaternionic projective representation of the translation group,
 introduced by Emch and Jadczyk,  and show in particular that this representation is strongly continuous. In Sec.~\ref{S4}, we define a quaternionic Weyl correspondence for the charged particle-monopole system,  using a quaternionic analog of the Fourier transform.  In Sec.~\ref{S5}, we prove a lemma on the Weyl symbols of operators of a special type, which  provides an optimal way of finding the star product generated by this correspondence. In Sec.~\ref{S6}, we use the Zassenhaus formula to write the multiplier of the above-mentioned quaternionic projective representation in the form that is most convenient for our purposes. In Sec.~\ref{S7}, we derive an exact expression for the integral kernel of  the emerging star product.
In Sec.~\ref{S8}, we present a simple algorithm for finding the differential form of this  product and calculate it explicitly up to the second order in $\hbar$.  Section~\ref{S9} contains concluding remarks and a comparison of the obtained star product   with the Kontsevich formula~\cite{K} for  deformation quantization of Poisson structures.

A few words about our notation concerning quaternions: The algebra of quaternions is denoted  $\mathbb H$ and its imaginary units are denoted $\mathfrak e_1$,$\mathfrak e_2$, and $\mathfrak e_3$.  The multiplication in $\mathbb H$ is defined by the relations
\begin{equation}
\mathfrak e_i\mathfrak e_j=-\delta_{ij}+\sum_k\epsilon_{ijk}\mathfrak e_k,
 \notag
\end{equation}
where  $\epsilon_{ijk}$ is the completely antisymmetric Levi-Civita symbol. For each quaternion $\mathfrak  a=a^0+a^1\mathfrak e_1+a^2\mathfrak e_2+a^3\mathfrak e_3\in \mathbb H$, its  conjugate   is defined as $\mathfrak  a^*=a^0-a^1\mathfrak e_1-a^2\mathfrak e_2-a^3\mathfrak e_3$ and the conjugation operation is an involution, i.e.,  $\mathfrak a^{**}=\mathfrak a$ and $(\mathfrak a\mathfrak b)^*=\mathfrak b^*\mathfrak a^*$.

\section{Magnetic Poisson brackets}\label{S2}

The motion of a  particle with charge $e$  and mass $m$ in a magnetic field $\mathbf B(x)$ is described by the equations
\begin{equation}
m\frac{d\mathbf v}{dt}= e\mathbf v\times \mathbf B,
 \label{2.1}
\end{equation}
where  $\mathbf v=\left(\dfrac{dx^1}{dt},\dfrac{dx^2}{dt},\dfrac{dx^3}{dt}\right)$  and the physical units are chosen so that the speed of light $c=1$. It is well known (see, e.g.,~\cite{MR}), that Eqs.~\eqref{2.1} can be rewritten in the Hamiltonian form
\begin{equation}
\dot x^i= \{x^i,H\}_B,\qquad \dot p_i=\{p_i, H\}_B
 \label{2.1*}
\end{equation}
by taking the kinetic energy as Hamiltonian,
\begin{equation}
H=\dfrac{1}{2m}\sum_ip^2_i,
\notag
\end{equation}
and defining the magnetic Poisson bracket by
\begin{equation}
\{f, g\}_{B}=\frac{\partial f}{\partial x^i}\frac{\partial g}{\partial p_i}-\frac{\partial f}{\partial p_i}\frac{\partial g}{\partial x^i}+\beta_{ij}\frac{\partial f}{\partial p_i}\frac{\partial g}{\partial p_j},\quad\text{where}\quad \beta_{ij}=e\epsilon_{ijk} B^k.
 \label{2.2}
\end{equation}
Then for the coordinate functions $x^i$ and $p_i$, we have the relations
\begin{equation}
\{x^i, x^j\}_{B}=0,\quad \{x^i, p_j\}_{B}=\delta^i_j,\quad  \{p_i, p_j\}_{B}=\beta_{ij}(x),
 \label{2.3}
\end{equation}
 and the matrix $\mathcal P$  of the Poisson structure has the form
\begin{equation}
\mathcal P=\begin{pmatrix}\beta(x) &-I_n\\I_n&0\end{pmatrix}
\label{2.4}
\end{equation}
if we assume that the momentum coordinates $p_i$  are placed first and the  position coordinates $x^i$ are placed second. Accordingly, the phase-space symplectic form is
\begin{equation}
\omega=dp_i\wedge dx^i+\frac{1}{2} \beta_{ij}\,dx^i \wedge dx^j.
 \label{2.5}
\end{equation}
(In~\eqref{2.2} and \eqref{2.5} and hereafter, we use the summation convention for repeated indices.)
If the magnetic field can be written as $\mathbf B=\curl\mathbf A$, i.e., $\beta_{ij}=e(\partial_iA_j-\partial_jA_i)$, where the magnetic vector potential  $\mathbf A$ is continuously differentiable, then Eq.~\eqref{2.1} is also Hamiltonian relative to
the standard symplectic structure  $dp_i\wedge dx^i$ with the Hamiltonian function $H_{\mathbf A}=\dfrac{1}{2m}(\mathbf p-e\mathbf A)^2$. But such a
representation is impossible globally in the case of the magnetic monopole field, which has the form
\begin{equation}
 B^k(x) = g\frac{x^k}{|x|^3},
 \label{2.6}
\end{equation}
 where $g$  is the monopole charge. The flux of the field~\eqref{2.6} through any sphere surrounding the monopole is  $4\pi g$, while the Stokes theorem gives zero for a field of the form  $\curl\mathbf A$. Therefore,  no smooth vector potential for the field~\eqref{2.6} exists on its domain of definition $\oR^3\setminus\{0\}$.
 Such a potential exists only on a smaller domain obtained by removing a curve that begins at the origin and proceeds to infinity in some direction, exhibiting itself as a line of singularity.

\section{A weak quaternionic projective representation of the translation group}\label{S3}

Let $L^2(\oR^3,\mathbb H)$  be the Hilbert space of quaternion-valued   functions on  $\oR^3$ with the inner product
 \begin{equation}
 \langle\Phi,\Psi\rangle=\int_{\oR^3}\! dx\, \Phi(x)^*\Psi(x).
 \notag
\end{equation}
Following Refs.~\cite{EJ,A1}, we assume that $L^2(\oR^3,\mathbb H)$  is a right module over  $\mathbb H$; i.e., the multiplication of vectors by quaternionic scalars is taken to act from the right, while linear  operators act from the left.  The Emch and Jadczyk approach~\cite{EJ} to the description of the quantum dynamics of a particle in the monopole field is based on using   anti-Hermitian operators $\nabla_i$ acting in  $L^2(\oR^3,\mathbb H)$ as follows:
\begin{equation}
 (\nabla_i\Psi)(x)=\left(\partial_i+\frac12\epsilon_{ijk}\frac{x^j}{|x|^2}\mathfrak e_k\right)\Psi(x).
 \label{3.2}
\end{equation}
It is easy to verify that they obey the commutation relations
\begin{equation}
 [\nabla_i,\nabla_j]=-\frac12 J\epsilon_{ijk}\frac{x^k}{|x|^3},
 \label{3.3}
\end{equation}
where $J$ is the operator of left multiplication by the $x$-dependent imaginary unit quaternion
\begin{equation}
 \mathfrak j(x)=\frac{x\cdot\mathfrak  e}{|x|}, \qquad \text{with}\quad x\cdot\mathfrak  e\eqdef x^i\mathfrak e_i.
 \label{3.4}
\end{equation}
The operators~\eqref{3.2} are defined by a natural   $SU(2)$ connection on the quaternionic line bundle  over $\oR^3\setminus\{0\}$ whose square-integrable sections form the space
 $L^2(\oR^3\setminus\{0\},\mathbb H)$ identical to  $L^2(\oR^3,\mathbb H)$ (for more details we refer the reader to Ref.~\cite{EJ} ). As we will soon see, these operators generate ``twisted'' translations. The operator $J$ satisfies the relations  $J^\dagger=-J$ and $J^\dagger J=JJ^\dagger=I$, i.e., is anti-Hermitian and unitary, and clearly $J^2=-I$. It is important that $J$ commutes with all the operators $\nabla_i$:
   \begin{equation}
 [J,\nabla_i]=0, \quad i=1,2,3.
 \label{3.5}
\end{equation}
 (But it should be noted that if  $\mathrm g\ne1$, then $J$ does not commute with the operators  $\partial_i+\dfrac {\mathrm g}{2}\epsilon_{ijk}\dfrac{x^j}{|x|^2}\mathfrak e_k$ considered in Ref.~\cite{CG-BLMV} along with $\nabla_i$.)  Clearly, $J$ commutes  with the position operators  $Q^i$ defined as usual, by
 \begin{equation}
 (Q^i\Psi)(x)=x^i\Psi(x).
 \label{3.6}
\end{equation}
  If the quantum Hamiltonian is taken to be  $\mathcal H=-\dfrac{\hbar^2}{2m}\nabla^2$ and the evolution operator is defined as   $\exp{\left(-\dfrac{J}{\hbar}\mathcal Ht\right)}$, then using the commutation relations~\eqref{3.3}, we obtain  the following evolution equations in the Heisenberg picture:
 \begin{gather}
 \dot{Q}_i=\frac{J}{\hbar}[\mathcal H,Q_i]=-\frac{J\hbar}{m}\nabla_i, \label{3.7} \\ \ddot{Q}_i=\frac{1}{2m}\cdot\frac{\hbar}{2g}\epsilon_{ijk}(\dot{Q}^jB^k+
 B^k\dot{Q}^j),
 \label{3.8}
\end{gather}
 where $B^k$ is the multiplication operator by the function~\eqref{2.6}.  Equations~\eqref{3.7} and  \eqref{3.8} correspond to the initial classical equations of motion~\eqref{2.1} and \eqref{2.1*} under the condition that the particle electric charge $e$ and the monopole magnetic charge $g$ are  connected by the relation
\begin{equation}
 eg=\frac{\hbar}{2}.
 \label{3.9}
\end{equation}

Now, let $a$ be a vector in $\oR^3$ and let $a\cdot\nabla= a^i\nabla_i$. To construct a quaternionic Weyl correspondence,  we  consider the family of operators
 \begin{equation}
 V(a)=e^{a\cdot\nabla},
 \label{3.10}
 \end{equation}
treating each one as
\begin{equation}
 V(a)=e^{a\cdot\partial}W(a),
 \label{3.11}
 \end{equation}
 where $a\cdot\partial=a^i\partial_i$. To find  $W(a)$, we introduce the operator function $W(a,t)=e^{-ta\cdot\partial} e^{ta\cdot\nabla}$ which becomes $W(a)$ at  $t=1$. Taking its derivative in respect to  $t$ and using that $e^{-ta\cdot\partial}$ is a shift operator, we obtain
\begin{equation}
 \frac{\partial}{\partial t} W(a,t)=\frac12\epsilon_{ijk}\frac{a^i(x-ta)^j}{|x-ta|^2}\mathfrak e_kW(a,t)=\frac12\,\frac{(a\times x)\cdot\mathfrak e}{|x-ta|^2}W(a,t).
 \label{3.12}
 \end{equation}
 The unique solution of Eq.~\eqref{3.12} subject to the initial  condition $W(a,0)=I$ is the operator of multiplication by the function
\begin{equation}
 \mathrm w(a,x,t)=\exp{\left(\frac12(a\times x)\cdot\mathfrak e\int_0^t \frac{ds}{|x-sa|^2}\right)}
 \notag
 \end{equation}
 which is well defined for any noncollinear vectors   $a$ and $x$. For $t=1$, the integral in the exponent equals  $\alpha(x,a)/|a\times x|$, where $\alpha$ is the angle between $x$ and $x-a$, i.e., $\cos\alpha= x\cdot (x-a)/|x||x-a|$, and we deduce that $W(a)$ is the unitary operator of multiplication by
\begin{equation}
 \mathrm w(a,x)=\exp{\left(\frac12\mathfrak j(a\times x)\alpha\right)}=\cos\frac{\alpha}{2}+ \mathfrak j(a\times x)\sin\frac{\alpha}{2}.
 \notag
 \end{equation}

As noted in Ref.~\cite{EJ}, the unitary operators~\eqref{3.11} define a weak projective representation of the  translation group  in the sense of Adler~\cite{A1,A2,AE}.
Namely, it follows from~\eqref{3.11} that
\begin{equation}
 V(a)V(b)=V(a+b)M(a,b),
 \label{3.15}
 \end{equation}
 where $M(a,b)$ is the operator of multiplication by
 \begin{equation}
 \mathfrak m(a,b;x)=\mathrm w(a+b,x)^*\mathrm w(a, x-b)\mathrm w(b,x) .
 \label{3.16}
 \end{equation}
 The word "weak"\, means that the quaternion unitary phase~\eqref{3.16}   has a nontrivial $x$-dependence.\footnote{Adler defined a weak projective group representation by $U_aU_b|f\rangle=U_{ab}|f\rangle\Omega(a,b;f)$, where $a$ and $b$ are group elements, $\{|f\rangle\}$ is a (privileged) complete set of states, and $\Omega(a,b;f)$ is a phase factor. There are subtleties when the states are unnormalized, but we use the name "weak projective representation"\, in this case, too, as it was used in Ref.~\cite{EJ}.}
 It is worth noting that $M(a,b)$ does not commute with $V(a+b)$  and is the right multiplier of this representation, whereas its left multiplier is  equal to $M(-b,-a)^\dagger$. The associativity relation  $(V(a)V(b))V(c)=V(a)(V(b)V(c))$ implies that $M(a,b)$ satisfies the  2-cocycle  condition
 \begin{equation}
 M(a+b,c)V(c)^{-1}M(a,b)V(c)=M(a, b+c)M(b,c),
 \label{3.17}
 \end{equation}
 where  $V(c)$ cannot be dropped because of the noncommutativity.
  If $a\to a_0$, then $\mathrm w(a,x)\to \mathrm w(a_0,x)$ almost everywhere and, since  $|\mathrm w(a,x)- \mathrm w(a_0,x)|\le2$, we find that
 \begin{equation}
 \|W(a)\Psi-W(a_0)\Psi\|^2_{L^2(\oR^3,\mathbb H)}\to 0
 \notag
 \end{equation}
by the Lebesgue dominant-convergence theorem.
Setting $\Psi_a=W(a)\Psi$ and writing
\begin{equation}
  V(a)\Psi- V(a_0)\Psi= e^{a\cdot\partial}(\Psi_a-\Psi_{a_0})+(e^{a\cdot\partial}-e^{a_0\cdot\partial})\Psi_{a_0},
 \notag
 \end{equation}
 we conclude that the unitary representation
 $a\to V(a)$ is strongly continuous; i.e., for each  $\Psi\in L^2(\oR^3,\mathbb H)$,
 \begin{equation}
  V(a)\Psi\to V(a_0)\Psi\quad \text{as}\quad  a\to a_0.
 \notag
 \end{equation}

\section{Quaternionic Weyl correspondence}\label{S4}

In this section, the phase-space coordinates are denoted  $(p,q)$, where $q=(q^1, q^2, q^3)$ is the position vector of a particle and $p=(p_1,p_2,p_3)$  is its momentum vector.
To construct a generalized Weyl correspondence for a charged particle in the monopole field, we consider an inclusion  $\mathfrak J$  of the set of complex phase-space functions into the set of quaternion-valued functions, which is defined by
\begin{equation}
 \mathfrak J\colon f(p,q) \longmapsto  \mathfrak f(p,q)\eqdef \Re f(p,q)+\mathfrak j(x)\Im f(p,q).
 \label{4.0}
\end{equation}
 We note an important difference between this definition and  formula  (5.1) of Ref.~\cite{CG-BLMV}, where $q=x$.  Although $\mathfrak f$ depends on $x$, hereafter we omit the argument $x$ for brevity, because this dependence is fixed. Clearly, the map~\eqref{4.0}  is injective, and it follows immediately from the equality $\mathfrak j(x)^2=-1$ that
 \begin{equation}
\mathfrak J(f+g)=\mathfrak f+\mathfrak g,\quad\mathfrak J((s+it)f)=(s+\mathfrak j(x)t)\mathfrak f\quad (s,t\in\oR),
 \label{4.1}
\end{equation}
 i.e., $\mathfrak J$ is a  semilinear  homomorphism\footnote{The set of quaternions of the form $s+\mathfrak j(x)t$ can be regarded as a set of scalars acting on the image of the map $\mathfrak J$.} and furthermore it preserves multiplication:
  \begin{equation}
\mathfrak J(fg)=\mathfrak f\,\mathfrak g.
 \label{4.2}
\end{equation}
 Assuming that $f$ has a  Fourier transform,   we define a ``quaternionic'' Fourier transform of $\mathfrak f$   by the relation
 \begin{equation}
\tilde{\mathfrak f}(u,v) =\frac{1}{(2\pi)^3}\int\! dpdq\, e^{-\mathfrak j(x)(p\cdot u+q\cdot v)}\mathfrak f(p,q).
 \label{4.3}
\end{equation}
 If $\mathfrak f$ depends only on  $p$, or only on   $q$, then its Fourier transform in this variable is also denoted $\tilde{\mathfrak f}$, when this  cannot cause confusion.
 Assuming that the complex-valued functions of $u,v$ are embedded in the space of quaternion-valued functions  in a manner analogous to~\eqref{4.0}, we have
 \begin{equation}
\mathfrak J(\tilde f)=\tilde{\mathfrak f},
 \label{4.4}
\end{equation}
  where $\tilde f$ is the ordinary Fourier transform. It is easy to see that familiar formulas of  Fourier analysis have analogs for the  transform~\eqref{4.3}. In particular,
 \begin{equation}
\frac{1}{(2\pi)^3}\int\!dp\, e^{-\mathfrak j(x)p\cdot u}=\delta(u)
 \notag
\end{equation}
and correspondingly
 \begin{equation}
\frac{1}{(2\pi)^3}\int\! dudv\, e^{\mathfrak j(x)(p\cdot u+q\cdot v)}\tilde{\mathfrak f}(u,v)=\mathfrak f(p,q).
 \notag
\end{equation}

 We let  $P_i$ denote the Hermitian momentum operators defined by
   \begin{equation}
 P_i=-\hbar J\nabla_i,
\label{4.7}
\end{equation}
 where $J$ is the operator of left multiplication by $\mathfrak j(x)$, as before. The operators $P_i$ and the position operators  $Q^i$ defined above by~\eqref{3.6} satisfy the commutation relations
  \begin{gather}
[Q^i,Q^j]=0,\quad [Q^i,P_j]=\hbar J\delta^i_j,\notag\\ [P_i,P_j]=\hbar J\beta_{ij},
 \label{4.8}
\end{gather}
where  $\beta_{ij}$ is the operator of multiplication by
\begin{equation}
\beta_{ij}(x)=
eg\varepsilon_{ijk}\frac{x^k}{|x|^3}
 \label{4.9}
\end{equation}
and where the charges obey the quantization condition~\eqref{3.9}.
Now we define a quaternionic Weyl correspondence associating operators on the quaternionic Hilbert space $L^2(\oR^3,\mathbb H)$  with complex-valued functions on the phase space by the rule  \begin{equation}
f \longmapsto  \mathcal O_f=\frac{1}{(2\pi)^3}\int\! du dv\,e^{J(u\cdot  P+v\cdot  Q)}\,\tilde{\mathfrak
f}(u,v),
 \label{4.10}
\end{equation}
 where the imaginary unit  $\mathfrak j(x)$ occurring in $\tilde{\mathfrak f}$ acts on  wave functions as $J$, i.e., by left multiplication. The map~\eqref{4.10}  is semilinear, as is the map~\eqref{4.0}.
 Following  the usual terminology of quantization theory, the function $f$ will be called the  Weyl symbol of the operator   $\mathcal O_f$. We note that
 \begin{equation}
\mathfrak J(\widetilde{p^\alpha q^\gamma)}= (2\pi)^3\mathfrak j(x)^{|\alpha+\gamma|}\partial^\alpha\delta(u)\partial^\gamma\delta(v),
 \notag
\end{equation}
 where we use the standard multi-index notation
 \begin{equation}
 \alpha=(\alpha_1,\alpha_2,\alpha_3),\quad |\alpha|=\alpha_1+\alpha_2+\alpha_3,\quad p^\alpha=p_1^{\alpha_1}p_2^{\alpha_2}p_3^{\alpha_3},\quad  \partial^\alpha_u=\partial^{\alpha_1}_{u_1}\partial^{\alpha_2}_{u_2} \partial^{\alpha_3}_{u_3}.
   \notag
\end{equation}
Therefore, the monomial $p^\alpha q^\gamma$ is transformed by the map~\eqref{4.10} into the symmetrically ordered product
  \begin{equation}
\{P^\alpha Q^\gamma\}_{\rm s}\eqdef\left.\frac{\partial^{\alpha+\gamma}e^{J(u\cdot P+v\cdot Q)}}{\partial u^\alpha\partial v^\gamma}\right|_{u,v=0} .
 \notag
\end{equation}
This is analogous to the usual Weyl correspondence for functions on a phase space with the standard symplectic structure, with  the difference that  in this case not only the operators $Q^i$ and $P_i$ but also the operators $P_i$ и $P_j$, where $i\ne j$, do not commute with each other. The star product $\star $ is by definition  the operation on the set of symbols which corresponds to the operator product under the map~\eqref{4.10}, and it easy to see that
 \begin{gather}
q^i\star q^j=q^iq^j,\quad q^i\star p_j= q^ip_j+ \frac{i\hbar}{2}\delta^i_j, \label{4.11}\\ p_i\star p_j= p_ip_j+\frac{i\hbar}{2}\beta_{ij}(q)
 \label{4.12}
\end{gather}
in accordance with the Poisson structure~\eqref{2.3}. By way of example, we prove the equality~\eqref{4.12}. As has just been said, the function $p_ip_j$ is transformed by the map ~\eqref{4.10} into the symmetric product
 \begin{equation}
\frac12(P_iP_j+P_jP_i).
 \label{4.13}
\end{equation}
Applying the Fourier transform~\eqref{4.3} to $\mathfrak J((i\hbar/2)\beta_{ij}(q))=(\hbar/2)\mathfrak j(x)\beta_{ij}(q)$,
we obtain
 \begin{equation}
\frac{1}{(2\pi)^3}\int\!dqdp\, e^{-\mathfrak j(x)(p\cdot u+q\cdot v)} \frac{\hbar}{2}\mathfrak j(x)\beta_{ij}(q)=
\frac{\hbar}{2}(2\pi)^{3/2}\mathfrak j(x)\delta(u)\tilde{\beta}_{ij}(v).
 \notag
\end{equation}
Substituting this expression  into~\eqref{4.10}  instead of $\tilde{\mathfrak f}(u,v)$ gives \begin{equation}
\frac{\hbar}{2}J\frac{1}{(2\pi)^{3/2}}\int\!dv\,
\tilde{\beta}_{ij}(v) e^{J v\cdot Q},
 \label{4.14}
\end{equation}
which is the multiplication operator by $(\hbar/2)\mathfrak j(x)\beta_{ij}(x)$ because  $(e^{J v\cdot Q}\Psi)(x)=e^{\mathfrak j(x)v\cdot x}\Psi(x)$ and the integration in~\eqref{4.14} yields the inverse Fourier transform.  Thus, on account of the commutation relation~\eqref{4.8}, the function $(i\hbar/2)\beta_{ij}(q)$ is transformed by ~\eqref{4.10}  into the operator $(1/2)[P_i,P_j]$, whose sum with ~\eqref{4.13} is $P_iP_j$, which proves the formula~\eqref{4.12}.

\section{An operator analog of the twisted convolution}\label{S5}

The unitary operators
 \begin{equation}
T(u,v)=e^{J(u\cdot  P+v\cdot  Q)}=e^{J u\cdot  P}e^{J v\cdot  Q}e^{-J\hbar  u\cdot v/2}
\label{5.1}
\end{equation}
  form a  weak quaternionic projective representation of the translation group of the phase space. Indeed, using~\eqref{3.15} and the commutation relation
 \begin{equation}
 e^{J v\cdot Q}e^{J u'\cdot P}=e^{J u'\cdot P}e^{J v\cdot Q}e^{-J\hbar u'\cdot v}
 \notag
\end{equation}
and letting $w$ denote for brevity the pair of variables $(u,v)$, we obtain
 \begin{multline}
T(w)T(w')=e^{J u\cdot  P}e^{J v\cdot  Q}
e^{J u'\cdot  P} e^{J v\cdot ' Q}e^{-J\hbar (u\cdot v+ u'\cdot v')/2}\\=
e^{J u\cdot  P}
e^{J u'\cdot  P}e^{J v\cdot  Q}e^{J v'\cdot  Q}e^{-J \hbar (u\cdot v+ u'\cdot v'+2u'\cdot v)/2}=
T(w+w')\mathcal M_\hbar(w,w'),
 \label{5.2}
\end{multline}
where   $\mathcal M_\hbar$ is the composite multiplier
\begin{equation}
\mathcal M_\hbar(w,w')= M(\hbar u,\hbar u')e^{J\hbar(u\cdot v'- v\cdot u')/2}
 \label{5.3}
\end{equation}
 and $M(\hbar u,\hbar u')$ is the operator of multiplication  by the quaternion-valued function  $\mathfrak m(\hbar u,\hbar u';x)$ defined by~\eqref{3.16}. We note that $\mathcal M_\hbar$  commutes with $J$ but does not commute with $T(w+w')$. It is natural to try to find the composition law for the quaternionic Fourier transforms that corresponds to the operator  multiplication, as this has been done by von Neumann~\cite{N}  for  the usual Weyl correspondence. It follows from the definition~\eqref{4.10} and from the relation~\eqref{5.2} that
 \begin{equation}
\mathcal O_f\mathcal O_g=\frac{1}{(2\pi)^6}\int\!dwdw'\, T(w+w')\mathcal M_\hbar(w,w')\tilde{\mathsf f}(w)\tilde{\mathsf g}(w').
 \notag
\end{equation}
Making the change of variables $w+w'\to w$, this can be written as
 \begin{equation}
\mathcal O_f\mathcal O_g=\frac{1}{(2\pi)^3}\int\! dw\, T(w) (\tilde{\mathsf f}\circledast_\hbar\tilde{\mathsf g})(w),
 \label{5.4}
\end{equation}
where
\begin{equation}
(\tilde{\mathsf f}\circledast_\hbar\tilde{\mathsf g})(w)= \frac{1}{(2\pi)^3}\int \! dw'\, \mathcal M_\hbar (w-w',w')
\tilde{\mathsf f}(w-w')\tilde{\mathsf g}(w').
 \label{5.5}
\end{equation}
In the case of the usual Weyl correspondence, where the multiplier of a complex projective representation of the translation group is  $\exp{(i\hbar(uv'- vu')/2)}$, an analogous expression is called the twisted convolution product~\cite{Mail,Fol,W} and the inverse Fourier transform converts it into the Moyal star product  $f\star_{\mathrm M}g$, i.e., into the phase-space function that corresponds to the operator  $\mathcal O_f\mathcal O_g$. The Moyal product has the form
 \begin{equation}
(f\star_{\mathrm M}g)(p,q)=f(p,q)\exp\left\{\frac{i\hbar}{2}
(\overleftarrow{\partial_q}\cdot
\overrightarrow{\partial_p}-\overleftarrow{\partial_p}\cdot
\overrightarrow{\partial_q})\right\}g(p,q)
 \notag
\end{equation}
and the bidifferential operator defining this product  is  obtained from the multiplier $\exp{(i\hbar(uv'- vu')/2)}$ by the simple replacements
 \begin{equation}
u\to -i\overleftarrow{\partial_p},\quad v\to -i \overleftarrow{\partial_q},\quad u'\to -i\overrightarrow{\partial_p},\quad v'\to -i\overrightarrow{\partial_q}.
 \label{5.6}
\end{equation}
This is easily verified by expanding the exponential  $\exp{(i\hbar((u-u')v'- (v-v')u')/2)}$ in a power series. Then the twisted convolution product becomes a sum of the usual convolution products of functions  obtained from  $\tilde f$ and $\tilde g$ by multiplication by some monomials in  $u$ and $v$.  Next, it suffices to take into account that  the Fourier transform converts the convolution product  into pointwise product  and converts multiplication by monomials into differentiation and vice versa.
The form of~\eqref{5.5} outwardly resembles that of the twisted convolution product, but an important difference is that this expression is not a complex-valued  but an operator-valued function  of $w=(u,v)$ because of the additional quaternionic factor   $\mathfrak m(\hbar u,\hbar u';x)$ depending on  $x$. This complicates finding the symbol of the operator $\mathcal O_f\mathcal O_g$ from~\eqref{5.5}.
Nevertheless, starting from this formula and using an expansion of the composite multiplier $\mathcal M_\hbar$ in powers of  $\hbar$, it is possible to obtain the star product generated by~\eqref{4.10},  as  will be shown in Sec.~\ref{S8}. As a first step in this direction, we calculate the symbol of the operator  $T(u,v)\mathcal C$,
where $\mathcal C$ if the multiplication operator by a function depending on $x$.

\begin{lemma}\label{L1}.  Let $c(x)$ be a complex-valued function and  $\mathcal C$ the operator of multiplication by  $\Re c(x)+\mathfrak j(x)\Im c(x)$ in the quaternionic Hilbert space $L^2(\oR^3,\mathbb H)$. Then the symbol of $T(u,v)\mathcal C$  under the correspondence~\eqref{4.10} is given by $f_{u,v}(p,q)=e^{i(u\cdot p+v\cdot q)}c(q+\hbar u/2)$.
\end{lemma}

{\it Proof.}
 Let $\mathfrak c(q)= (\mathfrak J c)(q)=\Re c(q)+\mathfrak j(x)\Im c(q)$ in accordance with the notation introduced in Sec.~\ref{S4}. By the definition~\eqref{4.3}, we have
  \begin{multline}
\tilde{\mathfrak f}_{u,v}(u',v') =\frac{1}{(2\pi)^3}\int \!dpdq \, e^{-\mathfrak j(x)(u'\cdot p+v'\cdot q)+\mathfrak j(x)(u\cdot p+v\cdot q)}\mathfrak c(q+\hbar u/2)\\=(2\pi)^{3/2}\delta(u'-u)\tilde{\mathfrak c}(v'-v)e^{\mathfrak j(x)\hbar u\cdot (v'-v)/2}.
 \label{5.7}
\end{multline}
Using~\eqref{5.1} and~\eqref{5.7}, we find that the operator corresponding to $f_{u,v}$ is given by
  \begin{multline}
\mathcal O_{f_{u,v}}=\frac{1}{(2\pi)^3}\int\! du' dv'\,e^{Ju'\cdot  P}e^{v'\cdot  Q} e^{-J\hbar u'\cdot v'/2}\,\tilde{\mathfrak
f}_{u,v}(u',v')\\=e^{Ju\cdot( P-\hbar v/2)}\frac{1}{(2\pi)^{3/2}}\int dv'e^{Jv'\cdot Q}\,\tilde{\mathfrak c}(v'-v).
 \notag
\end{multline}
Applying the operator $(2\pi)^{-3/2}\int dv'e^{Jv'\cdot Q}\,\tilde{\mathfrak c}(v'-v)$ to $\Psi\in L^2(\oR^3,\mathbb H)$ and integrating over  $v'$, we obtain
\begin{equation}
e^{\mathfrak j(x)v\cdot x}\left(\Re c(x)+\mathfrak j(x)\Im c(x)\right)\Psi(x)=\left(e^{Jv\cdot Q}\mathcal C\Psi\right)(x).
\notag
\end{equation}
 Hence, $\mathcal O_{f_{u,v}}=T(u,v)\mathcal C$,   which completes the proof.

 \section{The Zassenhaus  formula}\label{S6}

To derive the desired differential form of the star product, we will employ, instead of~\eqref{3.16}, another representation of the multiplier  $M(\hbar u,\hbar u')$, which can be obtained by using the  Baker-Campbell-Hausdorff formula for  products of exponentials of noncommuting variables $X$ and $Y$. More precisely, we prefer to use a modification of this formula which was proposed by Zassenhaus and is written as
\begin{equation}
 e^{X+Y}= e^{X}e^{Y}\prod_{n=2}^\infty e^{C_n(X,Y)}.
 \label{6.1}
 \end{equation}
  Here $C_n$ is a homogeneous Lie polynomial in   $X$ and $Y$ of degree $n$, i.e., a linear combination  of nested commutators of the form  $[Z_1,[Z_2,\dots,[Z_{m-1},Z_m]\dots]$, where each of $Z_i$ is $X$ or $Y$.
Written out explicitly, the first terms in~\eqref{6.1} are
 \begin{equation}
 C_2(X,Y)=-\frac12[X,Y],\quad C_3=\frac16[X,[X,Y]]+\frac13[Y,[X,Y]].
 \notag
 \end{equation}
 Several systematic  approaches to calculating the Zassenhaus terms $C_n$ have been carried out in the literature; for more details, see~\cite{CMN}, where an efficient recursive procedure expressing directly $C_n$ with
the minimum number of commutators required at each degree $n$ is  proposed.

In our case,  $X=Ju\cdot P$, $Y=Ju'\cdot P$ and it follows from~\eqref{4.8} that
\begin{equation}
 C_2(Ju\cdot P,Ju'\cdot P)=\frac{\hbar}{2} J u^i\beta_{ij}u'^j=
 \frac{\hbar}{2} J u\cdot \beta u'.
\label{6.2}
 \end{equation}
  Because $J$ commutes with $P_i$,  the calculation of $C_n$ for $n\ge 3$  reduces to differentiation of the functions $\beta_{ij}(x)$. In particular,
\begin{equation}
 C_3(Ju\cdot P,Ju'\cdot P)=- \frac{\hbar^2}{6}
J\left[u\cdot(u\cdot\partial)\beta u'+2 u\cdot(u'\cdot\partial)\beta u'\right].
 \label{6.3}
 \end{equation}
 Each  $C_n(Ju\cdot P,Ju'\cdot P)$ is the product of $J$ by a function of  $x$. Therefore, they commute with each other,  and the multiplier  $M(\hbar u,\hbar u')$ can be written as
\begin{equation}
M(\hbar u,\hbar u')=\exp{\left(-\sum_{n=2}^\infty C_n(Ju\cdot P,Ju'\cdot P)\right)}.
 \label{6.4}
 \end{equation}
 Using the explicit expressions given by~\eqref{6.2}  and~\eqref{6.3}, we  get the following representation for the quaternion-valued function $\mathfrak m(\hbar u,\hbar u';x)$ defining this operator:
 \begin{multline}
 \mathfrak m(\hbar u,\hbar u';x)= \exp\left(-\frac{\hbar}{2} \mathfrak j(x) u\cdot\beta(x)u'\right. \\  \left.+\frac{\hbar^2}{6}
\mathfrak j(x)\left[u\cdot(u\cdot\partial)\beta(x)u'+2 u\cdot(u'\cdot\partial)\beta(x)u'\right]\dots\right).
 \label{6.5}
\end{multline}

\section{The integral form of the star product}\label{S7}

 We let $m(\hbar u,\hbar u';x)$ denote the complex-valued function  obtained from $\mathfrak m(\hbar u,\hbar u';x)$ by replacing $\mathfrak j(x)$ with $i$. In other words,  $\Re m(\cdot\, ;x)$ is the scalar part of the quaternion  $\mathfrak m(\cdot\, ;x)$, and $\Im m(\cdot\, ;x)$ is the scalar part of $-\mathfrak j(x)\mathfrak m(\cdot\, ;x)$. Clearly,  $\mathfrak m(\cdot\, ;x)$ is in turn obtained from $m(\cdot\, ;q)$ by applying the mapping~\eqref{4.0} and subsequently setting $q=x$.
 For any fixed  $w=(u,v)$, it follows from the definition~\eqref{5.5} and  relations~\eqref{4.1}, \eqref{4.2},   and~\eqref{4.4} that the operator   $\mathcal C_w=(\tilde{\mathsf f}\circledast_\hbar\tilde{\mathsf g})(w)$  is the multiplication operator by the function  $\Re c_w(x)+\mathfrak j(x)\Im c_w(x)$, where
\begin{equation}
c_w(x)=\frac{1}{(2\pi)^3}\int du'dv' m(\hbar (u-u'),\hbar u';x)
e^{i\hbar(u\cdot v'- v\cdot u')/2} \tilde f(u-u',v-v')\tilde g(u',v').
\notag
\end{equation}
The formula~\eqref{5.4}  represents the operator    $\mathcal O_f\mathcal O_g$ as the integral  over   $w$ of an operator-valued function of the form considered in Lemma~\ref{L1} of Sec.~\ref{S5}.  From this lemma and the linearity of the map~\eqref{4.10}, we immediately deduce that the symbol of  $\mathcal O_f\mathcal O_g$, i.e., the star product$f\star_\hbar g$ can be written as
\begin{multline}
(f\star_\hbar g)(p,q)=\frac{1}{(2\pi)^6}\int dudv\,du'dv'\,e^{i(u\cdot p+v\cdot q)}   m(\hbar (u-u'),\hbar u';q+\hbar u/2)\\
\times e^{i\hbar(u\cdot v'- v\cdot u')/2}
\tilde f(u-u',v-v')\tilde g(u',v')
 \label{7.1}
\end{multline}
or, equivalently, as
\begin{multline}
(f\star_\hbar g)(p,q)=
  \frac{1}{(2\pi)^6}\int dudv\,du'dv'\,e^{i(u\cdot p+v\cdot q)}   m(\hbar u',\hbar(u- u');q+\hbar u/2)\\
\times e^{-i\hbar(u\cdot v'- v\cdot u')/2}
\tilde f(u',v')\tilde g(u-u',v-v').
 \label{7.2}
\end{multline}
Using~\eqref{5.1} and the formulas of Sec.~\ref{S4} for the quaternionic Fourier transform, it is easy to  verify directly  that the map~\eqref{4.10} transforms
the function~\eqref{7.1}  into the operator  $\mathcal O_f\mathcal O_g$ written in the form~\eqref{5.4}. An integral representation of the star product in terms of the functions $f$ and $g$ themselves can now readily be derived as follows. Making a change of integration variables in~\eqref{7.2}, we obtain
\begin{multline}
(f\star_\hbar g)(p,q)=\frac{1}{(2\pi)^6}\int du'dv'\,du''dv''e^{i(u'+u'')\cdot p+i(v'+v'')\cdot q}   m(\hbar u',\hbar u'';q+\hbar (u'+u'')/2)\\
\times e^{-i\hbar(u''\cdot v'- v''\cdot u')/2}
\tilde f(u',v')\tilde g(u'',v'')=\\=
\frac{1}{(2\pi)^6}\int du'du''dp'dp''e^{i(u'+u'')\cdot p-iu'\cdot p'-iu''\cdot p''} m(\hbar u',\hbar u'';q+\hbar (u'+u'')/2)\\
 \times f(p',q-\hbar u''/2) g(p'',q+\hbar u'/2).
 \notag
\end{multline}
Setting $q'=q-\hbar u''/2$ and $q''=q+\hbar u'/2$,  we arrive at the representation
  \begin{equation}
(f\star_\hbar g)(p,q)=\int dp'dq'\,dp''dq''\, K(p',q',p'',q'';p,q)f(p',q')g(p'',q''),
 \notag
\end{equation}
where the integral kernel is given by
\begin{multline}
 K(p',q',p'',q'';p,q)=\frac{1}{(\pi\hbar)^6}
 \exp\left\{-\frac{2i}{\hbar}[(p-p')(q-q'')-(p-p'')(q-q')]\right\}\\
 \times m(2(q''-q),2(q-q'); q-q'+q'').
  \label{7.3}
\end{multline}
The exponential in the first line of~\eqref{7.3} is the integral kernel of the Moyal star product, and the additional factor in the second line is  caused by the monopole magnetic field. If we keep only the first term  $C_2$ in the representation~\eqref{6.4} of the magnetic multiplier, dropping all other terms, then the corresponding "truncated"\, expression for $ m(2(q''-q),2(q-q'); q-q'+q'')$ is $\exp\{-q\cdot(q'\times q'')/| q-q'+q''|^3\}$ and, instead of the exact representation~\eqref{7.3} of the integral kernel, we obtain the approximate formula
\begin{equation}
 K_{\rm approx}=\frac{1}{(\pi\hbar)^6}
 \exp\left\{-\frac{2i}{\hbar}[(p-p')(q-q'')-(p-p'')(q-q')]-\frac{q\cdot(q'\times q'')}{| q-q'+q''|^3}\right\}.
   \notag
\end{equation}
The star product defined by $K_{\rm approx}$ is in agreement with the initial Poisson structure~\eqref{2.3}  at the first order in $\hbar$ but is not associative even  at the second order.

\section{The differential form of the star product}\label{S8}

Starting from the Zassenhaus formula~\eqref{6.5} and using  the Taylor series  expansion about the point  $q$, we can express the function $m(\hbar (u-u'),\hbar  u'; q+\hbar u/2)$ in~\eqref{7.1} as a series in powers of $\hbar$  with coefficient functions that are polynomials in the  entries of the  matrix $\beta(q)$, their partial derivatives at the point  $q$, and  the components of the vectors   $u-u'$  and $u'$. On substituting  this expansion   and the expansion  of  $\exp{(i\hbar((u-u')v'- (v-v')u')/2)}$ into~\eqref{7.1}, the right-hand side is expressed as the inverse Fourier transform in  $u$ and  $v$ of an infinite sum of terms of the form
\begin{equation}
\mathcal B(q)\frac{\hbar^n}{(2\pi)^3}\int \!
 du'dv'\, (u-u')^\alpha (v-v')^\gamma {u'}^{\alpha'} {v'}^{\gamma'} \, \tilde f(u-u',v-v')\,\tilde g(u',v'),
 \label{8.1}
\end{equation}
  where $\mathcal B(q)$ is  a monomial in  the matrix entries  $\beta_{ij}(q)$ and their partial derivatives, depending on the multi-indices  $\alpha$, $\gamma$, $\alpha'$ and  $\gamma'$. Taking the  inverse Fourier transform of~\eqref{8.1}  gives
  \begin{equation}
 \hbar^n(-i)^{|\alpha|+|\alpha'+|\gamma+|\gamma'|}\mathcal B(q)\,\partial^\alpha_p\partial^\gamma_qf(p,q)
\partial^{\alpha'}_p\partial^{\gamma'}_qg(p,q)
 \notag
\end{equation}
and we obtain a differential representation of the star product which has the form
  \begin{equation}
f\star_\hbar g= \sum_{n=0}^\infty \hbar^n  B_n(f,g)
 \label{8.2}
\end{equation}
with some bidifferential operators $B_n$. Going from the variables $u-u'$ and $u'$  to $u$ and $u'$, we conclude that the operators  $B_n$ in~\eqref{8.2} are obtained from the composite multiplier~\eqref{5.3} by the following simple algorithm:
\begin{enumerate}
\item
Expand the multiplier~\eqref{5.3}  in powers of $\hbar$ using the Zassenhaus formula
and replace  the imaginary unit quaternion $j(x)$ with the complex imaginary unit $i$.
\item
   Substitute  $q+\hbar(u+u')/2$ for the argument $x$ of the matrix-valued functions $\beta_{ij}(x)$ and  their partial derivatives occurring in this expansion, and then expand them in  Taylor series about the point $q$.
\item
    Combine like terms in the resulting expansion, and replace the variables $u$, $u'$, $v$, and $v'$ with differential operators by the rule~\eqref{5.6}.
\end{enumerate}

We illustrate this procedure by  finding explicit expressions for  $B_1$ and $B_2$.
It follows from~\eqref{6.5} that
\begin{multline}
m(\hbar u,\hbar u';x)= 1-\frac{i\hbar}{2} u\cdot\beta(x)u'-\frac{\hbar^2}{8} (u\cdot\beta(x)u')^2\\+\frac{i\hbar^2}{6}
\left[u\cdot(u\cdot\partial)\beta(x)u'+2 u\cdot(u'\cdot\partial)\beta(x)u'\right]+O(\hbar^3).
 \notag
\end{multline}
Using also the expansion
\begin{equation}
\exp\left\{\frac{i\hbar}{2}(u\cdot v'-v\cdot u')\right\}=1+
\frac{i\hbar}{2}(u\cdot v'-v\cdot u')
-\frac{\hbar^2}{8}(u\cdot v'-v\cdot u')^2+O(\hbar^3),
 \notag
\end{equation}
we find that
\begin{multline}
  m(\hbar u,\hbar u';x)\exp\left\{\frac{i\hbar}{2}(u\cdot v'-v\cdot u')\right\}= 1+\frac{i\hbar}{2}\left[u\cdot v'-v\cdot u'-u\cdot\beta(x)u'\right]\\-\frac{\hbar^2}{8}\left[(u\cdot v'-v\cdot u')^2-2(u\cdot v'-v\cdot u')(u\cdot\beta(x)u') +(u\cdot\beta(x)u')^2\right]\\+\frac{i\hbar^2}{6}
\left[u\cdot(u\cdot\partial)\beta(x)u'+2 u\cdot(u'\cdot\partial)\beta(x)u'\right]+O(\hbar^3).
 \label{8.3}
\end{multline}
To calculate the $\hbar^2$-order terms  of the star product, it suffices to substitute  $q$ for $x$ in the second and third lines of~\eqref{8.3} and  to retain the first-order term of the expansion
\begin{equation}
\beta(q+\hbar(u+u')/2)=\beta(q)+ (\hbar/2)((u+u')\cdot \partial)\beta\left|_q\right.+O(\hbar^2)
 \notag
\end{equation}
in the first line.  This yields two additional terms similar to those in the third line but with different coefficients. Combining the similar terms and making the replacements~\eqref{5.6}, we obtain the bidifferential operator
  \begin{multline}
1+i\frac{\hbar}{2}\left[\overleftarrow{\partial_q}\cdot \overrightarrow{\partial_p}-
\overleftarrow{\partial_p}\cdot\overrightarrow{\partial_q}
+\overleftarrow{\partial_p}
\cdot \beta\overrightarrow{\partial_p}\right]\\-\frac{\hbar^2}{8}
\left[(\overleftarrow{\partial_q}\cdot \overrightarrow{\partial_p}-
\overleftarrow{\partial_p}\cdot\overrightarrow{\partial_q})^2
+2(\overleftarrow{\partial_q}\cdot \overrightarrow{\partial_p}-\overleftarrow{\partial_p}\cdot
\overrightarrow{\partial_q})(\overleftarrow{\partial_p} \cdot \beta\overrightarrow{\partial_p})+(\overleftarrow{\partial_p}
\cdot \beta\overrightarrow{\partial_p})^2\right]\\
+\frac{\hbar^2}{12}\left[
\overleftarrow{\partial_p}\cdot(\overleftarrow{\partial_p}\cdot\partial_q)
\beta\overrightarrow{\partial_p}-
\overleftarrow{\partial_p}\cdot(\overrightarrow{\partial_p}\cdot\partial_q)
\beta\overrightarrow{\partial_p}\right] .
 \notag
\end{multline}
Hence, up to the second order in $\hbar$, the star product defined by the quaternionic Weyl correspondence~\eqref{4.10}  has the form
\begin{multline}
f\star_\hbar g=fg+i\frac{\hbar}{2}\left(\partial_{q^i}f \partial_{p_i}g-
\partial_{p_i}f  \partial_{q^i}g +\beta_{ij}\partial_{p_i}f
\partial_{p_j}g\right)\\
-\frac{\hbar^2}{8}
\left(\partial_{q^i}\partial_{q^j}f \partial_{p_i}\partial_{p_j}g-
2\partial_{q^i}\partial_{p_j}f \partial_{p_i}\partial_{q^j}g+\partial_{p_i}\partial_{p_j}f  \partial_{q^i}\partial_{q^j}g \right)\\
-\frac{\hbar^2}{4}
\beta_{ij}\left(\partial_{q^k}\partial_{p_i}f
\partial_{p_k}\partial_{p_j}g  -\partial_{p_k}\partial_{p_i}f
\partial_{q^k}\partial_{p_j}g\right) -\frac{\hbar^2}{8}\beta_{ij}\beta_{kl}\partial_{p_i}\partial_{p_k}f
\partial_{p_j}
\partial_{p_l}g\\
+\frac{\hbar^2}{12}\partial_{q^k}
\beta_{ij}\left(
\partial_{p_i}\partial_{p_k}f\partial_{p_j}g-
\partial_{p_i}f\partial_{p_k}\partial_{p_j}g\right) +O(\hbar^3).
 \label{8.4}
\end{multline}

\section{Conclusion}\label{S9}

The formula~\eqref{8.4}  is in  agreement with  Kontsevich's deformation quantization  formula~\cite{K} which gives an associative star product in the case of a nonconstant Poisson structure, i.e., when its defining matrix  $\mathcal P^{ab}$ depends nontrivially on the phase-space coordinates. Up to the second-order terms, Kontsevich's formula is written as\footnote{The expansion parameter denoted by   $\hbar$ in Ref.~\cite{K} corresponds to $i\hbar/2$ in our notation.}
\begin{multline}
f\star g=fg+i\frac{\hbar}{2} \mathcal P^{ab}\partial_a f
\partial_b g
-\frac{\hbar^2}{8}\mathcal P^{a_1b_1}\mathcal P^{a_2b_2}
\partial_{a_1}\partial_{a_2}f
\partial_{b_1}
\partial_{b_2}g\\
-\frac{\hbar^2}{12}\mathcal P^{a_1b_1} \partial_{b_1}
\mathcal P^{a_2b_2}(
\partial_{a_1}\partial_{a_2}f\partial_{b_2} g -
\partial_{a_2}f\partial_{a_1}\partial_{b_2}g) +O(\hbar^3).
 \label{9.1}
\end{multline}
We are dealing with the phase space $(\oR^3\setminus\{0\})\times\oR^3$, and it is easy to verify that if   $\mathcal P^{ab}$ has the block form~\eqref{2.4}, then the formula~\eqref{9.1} is equivalent to~\eqref{8.4}.
The associativity   up to the second order means that
\begin{equation}
(f\star g)\star h=f\star(g\star h)+O(\hbar^3)
 \notag
\end{equation}
or,  which is the same, that the following equalities hold:
\begin{gather}
B_1(f,g)h+B_1(fg,h)=fB_1(g,h)+B_1(f,gh), \notag\\
B_2(f,g)h+B_1(B_1(f,g),h) +B_2(fg,h)=fB_2(g,h)+B_1(f,B_1(g,h))+B_2(f,gh).
 \notag
\end{gather}
We note that these associativity conditions are fulfilled for the star product~\eqref{8.4} at any charges  $e$ and  $g$, although the  condition~\eqref{3.9} is crucial for constructing the quaternionic Weyl correspondence, i.e., for  operator quantization. We also remark that  in order to construct a quaternionic quantization map at  the condition  $eg=-\hbar/2$, a left Hilbert $\mathbb H$-module   could be used; i.e., the convention of left multiplication by scalars and right multiplication by operators could be adopted. Then the anti-Hermitian operators
\begin{equation}
 \overleftarrow{\nabla}_i=\partial_i-\frac12\epsilon_{ijk}\frac{x^j}{|x|^2}\mathfrak e_k
 \notag
\end{equation}
satisfy the commutation relations
\begin{equation}
 [\overleftarrow{\nabla}_i,\overleftarrow{\nabla}_j]=\frac12 \overleftarrow{J}\epsilon_{ijk}\frac{x^k}{|x|^3},
 \notag
\end{equation}
 where $\overleftarrow{J}$ is the operator of right multiplication by the imaginary unit quaternion~\eqref{3.4}, and this operator commutes with all $\overleftarrow{\nabla}_i$, $i=1,2,3$.

A generalized Weyl correspondence for the charged particle-magnetic monopole system  can, of course,  also  be constructed within the complex Hilbert space framework and at any integer  $n$ in the Dirac charge quantization condition $eg=n\hbar/2$, $n\in\oZ$.
This is done in Ref.~\cite{S16} on the basis a global Lagrangian description~\cite{S84} of this system as a constrained system with $U(1)$ gauge symmetry on the total space of the principal bundle  $\oC^2\setminus\{0\}\to(\oC^2\setminus\{0\})/U(1)= \oR^3\setminus\{0\}$  whose restriction to the unit sphere of $\oR^3$ is the Hopf bundle. The appropriate Hilbert space consists of the square-integrable functions on $\oC^2$ satisfying the equivariance condition  $\Psi(ze^{i\theta}, \bar ze^{-i\theta})=e^{-in\theta}\Psi(z, \bar z)$. A comparison of this and the quaternion-based approach to the magnetic monopole quantum mechanics will be given elsewhere.


\begin{thebibliography}{99}

\baselineskip=15pt

\bibitem{WY1} T.~T.~Wu and  C.~N.~Yang, {\it Concept of nonintegrable phase factors and global formulation of gauge fields}, Phys. Rev. D{\bf 12} (1975) 3845-3857.

\smallskip
\bibitem{WY2} T.~T.~Wu and  C.~N.~Yang, {\it Dirac's monopole without strings: Classical Lagrangian theory}, Phys. Rev. D{\bf 14} (1976) 437-445.

\smallskip
\bibitem{WY3} T.~T.~Wu and  C.~N.~Yang,  {\it Dirac monopole without strings: Monopole Harmonics}, Nucl. Phys.  B{\bf 107}  (1976) 365-380.

\smallskip
\bibitem{GP} W.~Greub and H.~R.~Petry, {\it Minimal coupling and complex line bundles}, J. Math. Phys. (N.Y.) {\bf 16} (1975) 1347-1351.

\smallskip
\bibitem{D1} P.~A.~M.~Dirac,  {\it Quantised singularities in the electromagnetic field}, Proc. R. Soc. London  A{\bf 133} (1931) 60-72.

\smallskip
\bibitem{D2}  P.~A.~M.~Dirac, {\it The theory of magnetic poles}, Phys. Rev. {\bf 74} (1948)  817-830.

\smallskip
 \bibitem{EJ} G.~Emch and A.~Z.~Jadczyk, {\it Weakly projective representations, quaternions and monopoles}, in  Stochastic Processes, Physics and Geometry: New Interplays, II. A Volume in Honor of Sergio Albeverio.  CMS Conference Proceedings Series, Vol.~{\bf 29}, (AMS, Providence, RI, 2000),  p.~157 [arXiv:quant-ph/9803002].

\smallskip
\bibitem{A1} S.~L.~Adler, {\it Quaternionic Quantum Mechanics and Quantum Fields} (Oxford Univ. Press, NY, 1995).

\smallskip
\bibitem{A2} S.~L.~Adler, {\it Projective group representations in quaternionic Hilbert space}, J. Math. Phys. (N.Y.) {\bf 37} (1996) 2352-2360 [arXiv:hep-th/9601047].

\smallskip
\bibitem{AE} S~.L.~Adler and G.~G.~Emch,
 {\it A rejoinder on quaternionic projective representations}, J. Math. Phys. (N.Y.) {\bf 38} (1997) 4758-4762 [arXiv:hep-th/9704121]

\smallskip
 \bibitem{Madore} J.~Madore, {\it Geometric methods in classical field Theory}, Phys. Rep. {\bf 75} (1981) 125-204.

\smallskip
\bibitem{CG-BLMV}   J.~F.~Carinena, J.~M.~Gracia-Bondia, F.~Lizzi, G.~Marmo, and  P.~Vitale,  {\it Star-product in the presence of a monopole}, Phys. Lett. A{\bf 374} (2010) 3614-3618 [arXiv:0912.2197].

\smallskip
\bibitem{K} M.~Kontsevich, {\it Defomation quantization of Poisson manifolds, I},  Lett. Math. Phys. {\bf 66} (2003) 157-216 [arXiv:q-alg/9709040].

\smallskip
\bibitem{Mail} J.~M.~Maillard, {\it  On the twisted convolution
product and the Weyl transformation of tempered distributions,} J. Geom. Phys. {\bf 3}  (1986) 231-261.

\smallskip
\bibitem{Fol} G.~B.~Folland, {\it Harmonic Analysis in Phase Space},
Annals of Math. Studies Vol.~{\bf 122} (Princeton Univ. Press,
Princeton, NJ, 1989).

\smallskip
\bibitem{W} M.~W.~Wong, {\it Weyl Transforms} (Springer, New York, 1998).

\smallskip
\bibitem{CMN} F.~Casas, A.~Murua, and M.~Nadinic, {\it  Efficient computation of the Zassenhaus formula}, Computer Phys. Commun. {\bf 183} (2012) 2386-2391 [arXiv:1204.0389].

\smallskip
\bibitem{M} M.~M\"uller, {\it Product rule for gauge invariant Weyl symbols and its application to the semiclassical description of guiding centre
motion,}  J. Phys. A: Math. Gen. {\bf 32} (1999) 1035–1052 [arXiv:quant-ph/9805025].

\smallskip
\bibitem{MP} M. M\u{a}ntoiu and R. Purice, {\it The magnetic Weyl calculus,} J. Math. Phys. (N.Y.) {\bf 45} (2004)  1394-1417.

\smallskip
\bibitem{MR}   J.~E.~Marsden and T.~S.~Ratiu, {\it Introduction to Mechanics and Symmetry: A Basic Exposition of Classical Mechanical Systems} (Springer, New York, 1999).

\smallskip
\bibitem{N} J. von Neumann, {\it Die Eindeutigkeit der Schr\"odingerschen Operatoren},  Mathematische Annalen {\bf 104} (1931), 570-578.

\smallskip
\bibitem{S16} M.~A.~Soloviev, {\it  Weyl correspondence for a charged particle in the field of a magnetic  monopole,} Theor. Math. Phys. {\bf 187} (2016) 782-795.

\smallskip
\bibitem{S84} M.~A.~Solov'ev, {\it Dirac monopole as a Lagrangian system in
a stratification space}, JETP Lett. {\bf 39} (1984) 714-716.


\end{thebibliography}
\end{document}